\documentstyle[12pt,aasms4]{article}

\begin{document}

\title{Long-term Variation of AGNs}

\author{J.H. Fan\altaffilmark{1,2,3}, G.Z. Xie\altaffilmark{4},
 G. Adam\altaffilmark{1}, S.L. Wen\altaffilmark{5}, Y. Copin\altaffilmark{1},
 R.G. Lin\altaffilmark{2}, J.M. Bai\altaffilmark{4} and Y.P. Qin\altaffilmark{4}}

\affil{ 1. CRAL Observatoire de Lyon,9, Avenue Charles André, 69 563 Saint-Genis-Laval Cedex, France \\  
2. Center for Astrophysics, Guangzhou Normal University, Guangzhou 510400, China\\ 3. Joint Laboratory for Optical Astronomy, Chinese Academy of Sciences, China\\ 4. Yunnan Observatory, Chinese Academy of Sciences, Kunming 650011, China\\ 5. Physics Department, Hunan Normal University, Changsha, China}

\begin{abstract}
In this paper, we will discuss the long-term variations in the optical and the 
infrared bands of some  AGNs. It is interesting to note that the reported 
periods of AGNs are of the similar values ($\sim$  10 years, see Fan et al. 
1998a, ApJ, in press, and references therein ).  DCF method shows the optical and infrared 
bands  are strongly correlated suggesting that 
the emission mechanisms in the two bands have a common origin.\\

\end{abstract}


\section{Introduction}

 The nature of active galactic nuclei (AGNs) is still an open problem. The
 study of AGN variability, such as periodicity analysis, of light curves, 
 can yield valuable information about their nature and the 
 implications for quasar modeling are extremely important. 
 Photometric observations of AGNs are important to construct their light 
 curves and to study their variation behavior on different time scales. 
 In this poster, we present both historical data and new observations for 
 PKS0735+178, OJ287, and BL Lacertae and discuss  long-term variations.

\section{Periodicity Analysis}
\subsection{PKS 0735+178}
 PKS 0735+178 has been observed for $\sim$ 90 years in the optical (Fan et al.
 1997, A\&AS, 125, 525; Ap\&SS, 249, 269 ) and $\sim$ 20 years in infrared 
 band (Lin \& Fan 1998, Ap\&SS, in press).  Largest amplitude of 
 $\Delta B=4^{m}.6$, $J = 2^m.47$, $\Delta H = 2^m.30$,
 and  $\Delta K = 2^m.40$ and  color index of $B-V =  0.55 \pm 0.16$; 
 $B-I = 1.55  \pm 0.22  $; $B-U =  0.58 \pm 0.10 $;  $B-R =  0.99\pm 0.15 $ 
 have been found. Two periodicity analysis including Jurkevich analysis show 
 that there is a 14.2 (or 28.69)-year period in the B light curve (see Fan et
  al. 1997a,b).  DCF analysis shows that the variations in the optical and
  near-IR are strongly correlated (Lin \& Fan 1998, Ap\&SS, in press).

\subsection{0851+202, OJ287}
 OJ 287 has been observed for more than 100 years in optical and two dozens 
 of years in infrared. The optical light curve shows a $\sim$12-year period 
 (Sillanpaa et al. 1988, ApJ, 325, 628). DCF analysis  shows that the opt/IR 
 varaitions are  strongly correlated, the variations ($\Delta U = 4^{m}.72$;
  $\Delta B = 5^{m}.93 $; $\Delta V = 5^{m}.18$; $\Delta R = 4^{m}.45$; 
 $\Delta I = 4^{m}.07$; $\Delta J = 3^{m}.87$; $\Delta H = 3^{m}.78$;
 $\Delta K = 3^{m}.54$)  and color index ( $U-B = -0.60 \pm 0.14 $; 
 $B-V =  0.46 \pm 0.17 $; 
 $B-I = 1.42  \pm 0.25  $; $V-R =  0.48 \pm 0.14  $; 
 $V-I =  0.98 \pm 0.22  $; $R-I =  0.56 \pm 0.12  $; 
 $J-H =  0.82 \pm 0.16  $; $H-K =  0.88 \pm 0.14  $;
  $J-K =  1.70 \pm 0.19 $) are found (see Fan et al. 1998b, A\&AS, in press). 
 The largest  amplitude variations give 
 $\Delta m_{\lambda}=-(2.24\pm0.29)log\lambda + (4.20\pm0.002)$

\subsection{BL Lacertae}
 BL Lacertae has been observed for more than 100 years in the optical 
 (Fan et al. 1998a) and $\sim$20 years in IR (Fan et al. 1998c, A\&AS, 
 in press). Variations  of  $\Delta U = 5^m.12$,  $\Delta B = 5^m.31$,
 $\Delta V = 4^m.73$, 
 $\Delta R = 2^m.73$,  $\Delta I = 2^m.54$,  $\Delta J = 2^m.29$,  
 $\Delta H = 2^m.42$,  $\Delta K = 2^m.93$ are found from the light curves. 
 A 14-year period is found in the B light curve (Fan et al. 1998a).

\section{Infrared variation}

 BL Lacertae objects show variations at all frequencies, the variations in 
 the infrared are as comparable as those  in the optical
 for well observed objects. It is common for J-K to be correlated with J-H
  and H-K but not for J-H and H-K.  
 From the compilation, we found that the positive correlation between color 
 index and magnitude do not always hold. Some
 objects (0138-097, 0735+178, 0537-441, 0851+202, 1215+303, 1219+285, 
 1749+096, 0215+015) show a positive correlation  between magnitude and 
 color index suggesting that the spectrum flattens 
 when the source brightens while others (0754+100, 1147+245, 1418+546, 
 1727+502) show an opposite behaviour. 

The averaged color index and redshift are found positively correlated.

\acknowledgments
JHF thanks Prof. Roland Bacon, the director of observatoire de Lyon to 
support him to attend the conference, he is also grateful to the
Education Fund Committee of Guangzhou city for support. This work is 
supported by the National Scientific Foundation of China (the Ninth Five-year 
Important Project) and the National Pandeng Project of China

\begin{figure}
\epsfxsize=15cm
$$
\epsfbox{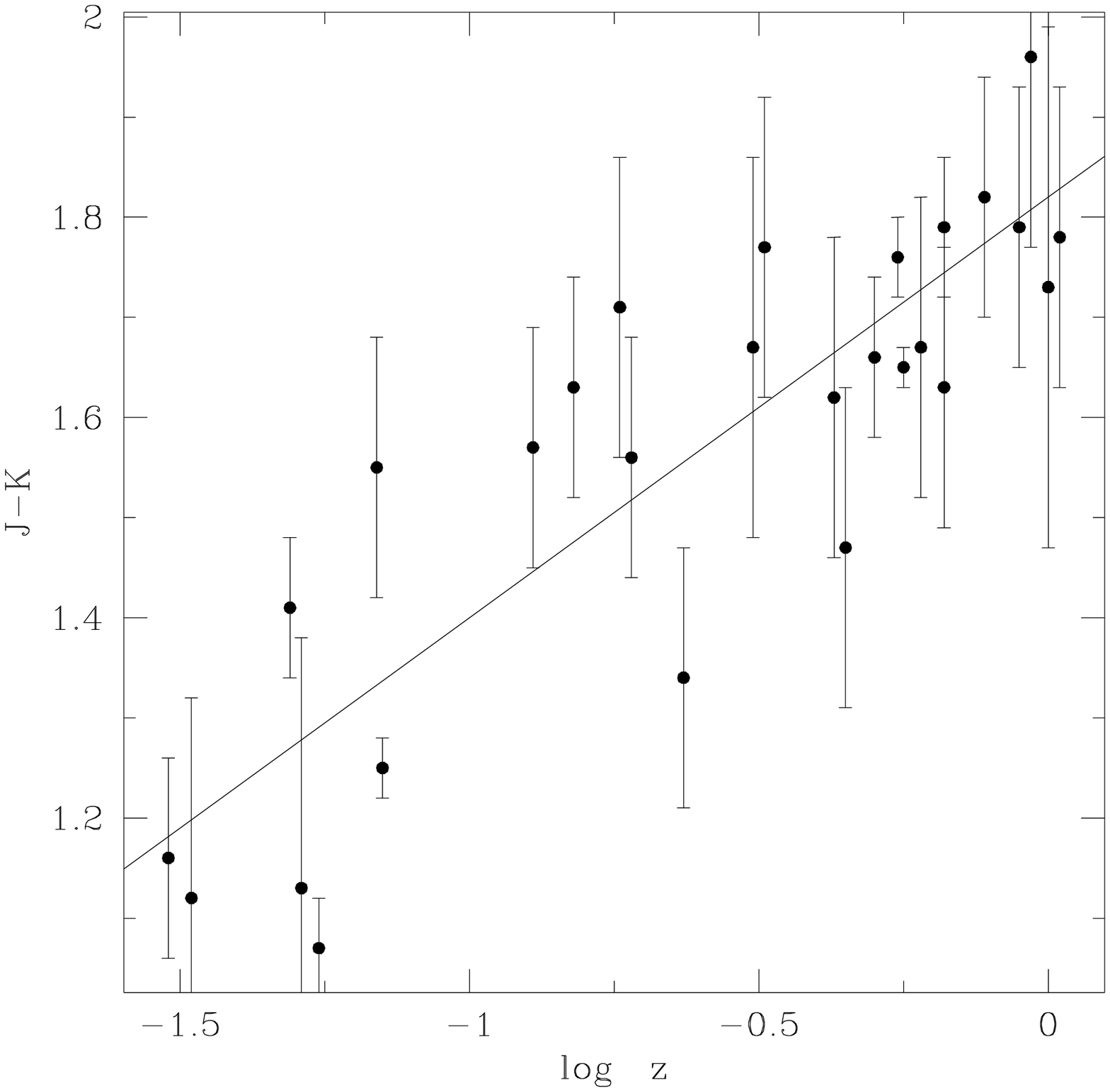}
$$

\caption{ Relation between color index and redshift for RBLs}
\end{figure}

\end{document}